\documentclass[12pt] {article}
\usepackage{latexsym}
\newcommand{\beq}{\begin{equation}}
\newcommand{\feq}[1]{\label{#1} \end{equation}}
\newcommand{\beqr}{\begin{eqnarray}}
\newcommand{\feqr}{\end{eqnarray}}
\def\non{\nonumber}
\def\noi{\noindent}
\newcommand{\rf}[1]{~(\ref{#1})}
\def\pr{^{\prime}}
\def\np#1#2#3{Nucl. Phys. {\bf{B#1}} (#2) #3}

\def\pr#1#2#3{Phys. Rep. {\bf{#1}} (#2) #3}
\def\prev#1#2#3{Phys. Rev. {\bf{D#1}} (#2) #3}

\def\cqg#1#2#3{Class. Quantum Grav. {\bf{#1}} (#2) #3}
\def\rmp#1#2#3{Rev. Mod. Phys. {\bf{#1}} (#2) #3}
\def\ptr#1#2#3{Prog. Theor. Phys. {\bf{#1}} (#2) #3}
\def\ap#1#2#3{Ann. of Phys. {\bf{#1}} (#2) #3}
\def\pla#1#2#3{Phys. Lett. {\bf{A#1}} (#2) #3}
\def\plb#1#2#3{Phys. Lett. {\bf{B#1}} (#2) #3}
\def\mpl#1#2#3{Mod. Phys. Lett. {\bf{A#1}} (#2) #3}
\def\jmp#1#2#3{J. Math. Phys. {\bf{#1}} (#2) #3}
\begin{document}

\renewcommand{\thefootnote}{\fnsymbol{footnote}}

\begin{center}

\vspace{2cm}

{\LARGE \bf Locally Weyl invariant massless bosonic and fermionic spin-1/2 action in the 
$\bf (W_{n(4)},g)$ and $\bf (U_{4},g)$ space-times.}\\[6mm]

\large{Agapitos Hatzinikitas}\footnote [2] 
{e-mail: ahatzini@cc.uoa.gr}\\[8mm]

{\it University of Athens, \\
Nuclear and Particle Physics Division,\\
Panepistimioupoli GR-15771 Athens, Greece.}\\[8mm]

{\small \bf Abstract}\\[3mm]

\parbox{6.2in}
{\small We search for a real bosonic and fermionic action in four dimensions which both remain invariant 
under local Weyl transformations in the presence of non-metricity and contortion tensor. In the presence of
the non-metricity tensor the investigation is extended to Weyl $(W_n, g)$ space-time while when the torsion is 
encountered we are restricted to the Riemann-Cartan $(U_4, g)$ space-time. Our results hold for a subgroup of the 
Weyl-Cartan $(Y_4, g)$ space-time and we also calculate extra contributions to the conformal gravity.}

\end{center}


%\newpage

%%%%%%%%%%%%%%%%%%%%%%%%%%%%%%%%%%%%%%%%%%%%%%%%%%%%%%%%%%%%%%%%%%%%%%%%%%%%%%%%%%%%%%%%%%%%%%%%%%%%%%%%%%%%%%%%%

					%%%%%%%%%%%%%%%%%%
					%  INTRODUCTION  %
					%%%%%%%%%%%%%%%%%%
\section{Introduction}

\renewcommand{\thefootnote}{\arabic{footnote}}
\setcounter{footnote}{0}

The notion of conformal invariance of microscopic phenomena is among the oldest and most intriguing ones
in modern theoretical physics. Physically conformal invariance means that nothing depends on the choice of 
dimensional units such as lengths, etc. Conformally invariant field theories reveal some advantages 
ranging from classical equations of motion up to better quantum behaviour (the
renormalization properties of matrix elements of the energy-momentum tensor are improved \cite{roman}, quantum gravity
is aymptotically free when a renormalization group analysis is performed \cite{fradkin}). 
 Realistic gravity or unified field theory probably cannot be conformally invariant but many 
theories can stem from the spontaneously broken versions of invariant models. 

In this work the non-metricity and torsion tensor are taken into account. The vanishing of torsion for the  
real world was Einstein's point of view that prevailed for a long time. This ``superfluous''
restriction seemed to add extra complications to the theory of General Relativity which turned out to have enormous 
success. On the other hand there was no compelling experimental reason to relax this condition. 

In the case of Weyl's geometry one may achieve to make contact with Einstein gravity by resorting either to gauge fixing 
or to a dynamical symmetry breaking mechanism. The reduced space is Riemannian and the Weyl vector propagates as a 
massive spin-1 particle mediating the neutrino-neutrino interaction \cite{hochberg}.  

The main theoretical 
advantage of gravity with torsion is that it links the spin of the matter fields with the space-time geometry 
\cite{gravto}. Thus classical particles in the presence of torsion will follow different geodesics but there is 
no experimental measure to confirm such an effect. Phenomenological aspects of the torsion in the context of Standard 
Model were studied in \cite{shapiro}. These authors by implementing the propagating 
torsion into the abelian sector of the Standard Model were able to extract information about the torsion mass 
and torsion-fermion coupling by studying the four fermion contact interactions. 

One could also be introduced to the exotic 
notion of torsion by the nonlinear sigma model action where it is represented locally on the field manifold as the 
curl of a second rank antisymmetric $B_{\mu \nu}$ \cite{zachos} potential according to:
\beqr
S_{\mu \nu \lambda}=\partial_{[ \mu}B_{\nu \lambda ]}.
\label{curza}
\feqr

The present paper attempts to produce concrete expressions for conformally invariant, massless bosononic and fermionic 
spin-$1/2$ actions equipped with a general affine connection. To fulfil this task we organize our work in the following 
way. 

Section 2 is devoted to a brief review of the background notions of the gravity 
with non-metricity and torsion tensor. Our main concern is the decomposition of the contortion into 
irreducible representations of the proper orthochronous Lorentz group. This spliting will be proved to be extremely 
useful in the construction of the fermionic action. Also we state the well-known way the Christoffel symbol as well as 
the spin connection transform under local Weyl transformations. 

In section 3 starting from the Einstein-Hilbert action 
with a cosmological constant in n-dimensions and performing Weyl transformations to the determinant and the Ricci 
scalar of the space-time we end up with a massless bosonic action. We then proceed to incorporate initially the 
non-metricity tensor by constructing a conformally invariant action in $(W_n, g)$ space-time. 
To achieve this, a suitably defined conformally 
covariant derivative is needed. When we encounter the torsion it 
is possible to build an affine connection that leads to a conformally invariant action. 

In section 4 using the decomposition of contortion, 
mentioned in section 2, we write down the corresponding fermionic spin-1/2 conformal action  
for the two distinct cases of non-metricity  
and torsion tensor, in $(W_4, g)$ and $(U_4, g)$ space-times respectively. In both cases one can define once 
more conformally 
covariant derivates for fermions and convince itself that the tensor part of the torsion decouples from the fermionic 
action. 

Section 5 enumerates and comments on which additional terms might contribute to the action. The selection is based on 
the conformal invariance of the free field  theory and the restriction to quadratic fields and derivatives 
of $\mathcal{R}$'s.These terms such as the Gauss-Bonnet theorem, the Pontrjagin and winding numbers are of 
topological nature. The Weyl tensor is also calculated and permits one to identify the terms introduced in the 
conformal gravity action in a rather natural way avoiding insertions by hand.

Finally, in the Appendix we establish all the necessary notation and identities employed in this paper. 
In the $(Y_4, g)$ space-time we write down the expressions giving the Riemann curvature, Ricci tensor and scalar 
as well as the corresponding quantities for the $(W_n, g)$ general case.

%%%%%%%%%%%%%%%%%%%%%%%%%%%%%%%%%%%%%%%%%%%%%%%%%%%%%%%%%%%%%%%%%%%%%%%%%%%%%%%%%%%%%%%%%%%%%%%%%%%%%%%%%%%%

					%%%%%%%%%%%%%%%%%%
					%   SECTION 1    %
					%%%%%%%%%%%%%%%%%%

\section{Background aspects of the gravity with torsion and non-metricity tensor.}

Any known conformally invariant field theory consists of two basic ingredients (see the review articles \cite{hehl}
and the huge amount of references therein). The first is the conformal transformation of the metric:
\beqr
\hat{g}_{\mu \nu}(x)=\Omega (x) g_{\mu \nu}(x),
\label{conmet}
\feqr

\noi and the second the affine connection $\tilde{\Gamma}^{\mu}_{\,\, \nu \lambda}(g, T, M)$ which is 
defined by:
\beqr
\nabla_{\mu}g_{\nu \lambda}=N_{\mu \nu \lambda}.
\label{metcomp}
\feqr

\noi As an exercise one can show that\rf{metcomp} leads to the most general form of the affine connection 
which reads \footnote{Tilded and hatted tensors are refered to $(Y_{4},g)$ space-time 
and their conformal equivalents respectively. The Riemann-Cartan $U_{4}$ space-time is a paracompact, Hausdorff, 
connected $C^{\infty}$
four dimensional manifold endowed with a locally Lorentzian metric and a linear affine connection obeying the 
metric compatibility condition: $\nabla_{\mu} g_{\nu \lambda}=\partial_{\mu}g_{\nu \lambda} - 
\Gamma^{\rho}_{\nu \mu} g_{\rho \lambda} - \Gamma^{\rho}_{\lambda \mu}g_{\rho \nu}=0$. 
In the present paper we will relax this restrictive condition to: $\nabla_{\mu} g_{\nu \lambda}=N_{\mu \nu \lambda}$ 
for which $ds^2=0$ is not preserved. One can require that $L_{\mu \nu \lambda}=L_{\mu (\nu \lambda)}$ and 
$g^{\nu \lambda}L_{\mu \nu \lambda}=0$.
(instead one could equally make $L_{\mu \nu \lambda}$ completely symmetric). $L_{\mu \nu \lambda}$ is called the shear 
tensor since it stretches and shrinks length.}:
\beq
\tilde{\Gamma}^{\mu}_{\,\,\, \nu \lambda}(g, T, M)=\Gamma^{\mu}_{\,\,\, \nu \lambda}(g)+ A^{\mu}_{\,\,\, \nu \lambda}
(T, M).
\feq{affgen}

\noi $\Gamma^{\mu}_{\,\,\, \nu \lambda}(g)=\{ {\mu \atop \nu \lambda} \}$ is the usual Christoffel symbol 
in the Riemann space-time
$V_{4}$, $A^{\mu}_{\,\,\, \nu \lambda}=K^{\mu}_{\,\,\, \nu \lambda} + M^{\mu}_{\,\,\, \nu \lambda}$ with 
$K^{\mu}_{\,\,\, \nu \lambda}$ the so-called contortion tensor defined by \footnote{Dots over the indices in\rf{condf} 
are reserved to keep track of their antisymmetry when needed.}:
\beqr
K_{\mu \nu \lambda} \!\!\! &=& \!\!\! \frac{1}{2} \left(T^{\,\,\,\, \cdot \, \cdot}_{\mu \nu \lambda} - 
T^{\,\,\,\, \cdot \, \cdot}_{\nu \mu \lambda} -
T^{\,\,\,\, \cdot \, \cdot}_{\lambda \mu \nu}  \right) \non \\
\!\!\! &=&  \!\!\! \frac{1}{2} \left(T^{\,\,\,\, \cdot \, \cdot}_{\mu \nu \lambda} + 
T^{\,\,\,\, \cdot \, \cdot}_{\nu \lambda \mu} +
T^{\,\,\,\, \cdot \, \cdot}_{\lambda \nu \mu}  \right)
\label{condf}
\feqr

\noi and
\beqr
M_{\mu \nu \lambda}=\frac{1}{2} \left( N_{\mu \nu \lambda} - N_{\nu \mu \lambda}
- N_{\lambda \mu \nu}\right)
\label{mdef}
\feqr

\noi where $N_{\mu \nu \lambda}=w_{\mu} g_{\nu \lambda} + L_{\mu \nu \lambda}$ and 
the Cartan's torsion is defined in terms of the antisymmetric part of the affine connection as:
\beq
T^{\mu \, \cdot \cdot}_{\,\,\,\, \nu \lambda}=2\tilde{\Gamma}^{\mu}_{\,\,\, [\nu \lambda]}
\feq{deft}

\noi In contrast to $\Gamma^{\mu}_{\nu \lambda}(g)$, the torsion is a proper tensor under general coordinate transformations.

The tensor $K_{\mu \nu \lambda}$ has the following properties:
\beqr
K_{\mu \nu \lambda} &=& - K_{\nu \mu \lambda} \\
K^{\mu}_{\,\,\, \mu \nu} = 0; \quad K_{\nu \mu}^{\ \  \mu} &=& -K^{\mu}_{\,\, \nu \mu}=v_{\nu}; \quad 
K_{\mu \nu \lambda} = \frac{1}{2} T^{\, \cdot \, \cdot \, \cdot}_{\mu \nu \lambda}
\label{Kpro}
\feqr 

\noi while $M_{\mu \nu \lambda}$ satisfies:
\beqr
M_{\mu \nu \lambda} &=& M_{\mu (\nu \lambda)} \\
M^{\mu}_{\,\,\, \mu \nu} &=& - 2 w_{\nu} \\
M_{\mu \nu}^{\ \ \ \nu} &=& w_{\mu}-L_{\nu \mu}^{\ \ \ \nu} 
\label{Mprop}
\feqr 

\noi In\rf{condf} the last two 
terms belong, among others, to the  symmetric part of the affine connection and the last property in\rf{Kpro} holds 
on the condition that the torsion is totally antisymmetric. 

In four space-time dimensions the $24$ independent components of torsion can be covariantly split into a traceless part and 
a trace: $T_{\mu \nu \lambda}=Z_{\mu \nu \lambda}+\frac{2}{3} g_{\mu[\nu}v_{\lambda]}$. 
In particular the decomposition into irreducible representations of 
the proper orthochronous Lorentz group $SO(3,1)$ is as follows \cite{kenji,mccrea}:

\begin{enumerate}

\item the vector part $v_{\lambda}= g^{\mu \nu} T_{\mu \nu \lambda}$ of dimension $4$ 
transforming as the $(1/2,1/2)$ \footnote{The universal covering, i.e. the double covering, of $SO(3,1)$ 
is $SL(2,C)$ and its irreducible 
representations are parametrized by $(j_{1},j_{2})$ where ${\bf J}^{2}_{(i)} \rightarrow j_{i}(j_{i}+1)$, $i=1,2$
\cite{djord}.},

\item the axial-vector part: $S_{\mu}=\frac{1}{6}\epsilon_{\mu \nu \lambda \rho} T^{\nu \lambda \rho}$,
 (or parity violating term) of the same dimension as before and corresponding to the Young tableau 
$[111]$ and

\item the tensor $t_{\mu \nu \lambda}$ of mixed symmetry and dimension $16$ \footnote{In the metric-affine 
space $(L_{d},g)$ the irreducible representations have dimensions: $d$, $\frac{1}{6}d(d-1)(d-2)$ 
and $\frac{1}{3}d(d^{2} -4)$ respectively.} transforming according to $(3/2,1/2)$ $ \oplus (1/2,3/2)$ and 
associated with the Young tableau $[21]$. It is expressed by:
\beq
t_{\mu \nu \lambda}=T_{(\mu \nu) \lambda}-\frac{1}{3} g_{\mu \nu} v_{\lambda}+\frac{1}{3} 
g_{\lambda (\mu} v_{\nu)}.
\feq{tensor}

\noi The tensorial part satisfies the following properties:

\begin{enumerate}

\item is symmetric w.r.t. the first two indices:
\beqr
t_{\mu \nu \lambda}=t_{(\mu \nu) \lambda},
\label{sym}
\feqr

\item remains invariant under cyclic permutation of the indices:
\beqr
t_{\mu \nu \lambda}+t_{\lambda \mu \nu}+t_{\nu \lambda \mu}=0,
\label{cycl}
\feqr

\item and is traceless in the sense:
\beqr
g^{\mu \nu}t_{\mu \nu \lambda}= g^{\nu \lambda} t_{\mu \nu \lambda} = 
\epsilon^{\mu \nu \lambda \rho} t_{\nu \lambda \rho}=0.
\label{trace}
\feqr

\end{enumerate}

\end{enumerate}  

\noi Properties\rf{sym} and\rf{trace} are direct consequences of\rf{tensor} while\rf{cycl} holds in general for any Young
tableau with the structure $[21]$. The torsion and contortion can be reexpressed in terms of the irreducible 
representations as:
\beqr
T_{\mu \nu \lambda}=\frac{4}{3}t_{\mu [\nu \lambda]}+\frac{2}{3} g_{\mu [\nu }v_{\lambda]}+
\epsilon_{\mu \nu \lambda \rho}S^{\rho},
\label{torirr}
\feqr

\beqr
K_{\mu \nu \lambda}=-\frac{4}{3} t_{\lambda  [\mu \nu]} + \frac{2}{3} 
g_{\lambda [\nu }v_{\mu ]} + \frac{1}{2} 
\epsilon_{\mu \nu \lambda \rho} S^{\rho}.
\label{conirr}
\feqr

The conformal group $C^{\infty}_{\Omega}$ is infinite dimensional and the transformations
of tensors (denoted by capital letters) for the subgroup of dilatations fall into two classes:
\beqr
\textit{Class-I:} \quad A \rightarrow \hat{A} = \Omega^{d(A)} A
\label{clacon1}
\feqr

\beqr
\textit{Class-II:} \quad B \rightarrow \hat{B} = F(\Omega^{d(B)},\partial \ln \Omega),
\label{clacon2}
\feqr

\noi where $d(A)$ is the conformal weight (a real constant) of the quantity $A$ with suppressed indices. 
In the $V_{4}$ space-time the transformations:
\beqr
g_{\mu \nu} \!&\rightarrow&\! \hat{g}_{\mu \nu} =  \Omega g_{\mu \nu} \\
e^{a}_{\mu} \!&\rightarrow&\! \hat{e}^{a}_{\mu} = \Omega^{1/2} e^{a}_{\mu} 
\label{v4tra}
\feqr 

\noi determine uniquely the transformation law of the Christoffel symbol to be:
\beq
\Gamma^{\mu}_{\,\,\, \nu \lambda}\rightarrow \hat{\Gamma}^{\mu}_{\,\,\,\nu \lambda}=\Gamma^{\mu}_{\,\,\, \nu \lambda}+
\frac{1}{2} \left(\delta^{\mu}_{\nu} \, \partial_{\lambda} \! \ln \Omega +
\delta^{\mu}_{\lambda} \, \partial_{\nu} \ln \! \Omega -
g_{\nu \lambda} \, \partial^{\mu} \ln \! \Omega\right)
\feq{chrtr}

\noi where $\Omega(x)$ is a positively defined function. Both the explicit form and the transformation law of 
the spin connection are then deduced from the vierbein postulate:
\beqr
D_{\mu}(\Gamma,\omega)e^{a}_{\nu}(x)=
\partial_{\mu}e^{a}_{\nu}-\Gamma^{\lambda}_{\,\, \nu\mu}e^{a}_{\lambda}+
\omega_{\mu \,\,\, b}^{\,\,\, a} \, e^{b}_{\nu}=0
\label{vilbr}
\feqr 

\noi and are:
\beqr
\omega_{\mu a b} &=& e^{\nu}_{b} \partial_{[ \nu} e_{\mu ] a} + e^{\rho}_{a} \partial_{[ \mu}e_{\rho ]b}
+ e^{\rho}_{a} e^{\nu}_{b} e^{j}_{\mu} \partial_{[\nu} e_{\rho ]j} \non \\
&=& \frac{1}{2} \left(\lambda_{b \mu a}+ \lambda_{\mu a b}+ \lambda_{ba \mu} \right)
\label{explco}
\feqr

\beqr
\omega^{\,\,\, a}_{\mu \,\,\,\, b} \rightarrow \hat{\omega}^{\,\,\, a}_{\mu \,\,\,\, b}=
\omega^{\,\,\, a}_{\mu \,\,\,\, b}+ \frac{1}{2} e^{a}_{\lambda}
\left( e^{\nu}_{b} \, \delta^{\lambda}_{\mu} \, \partial_{\nu} \! \ln \Omega -
e_{\mu b} \, \partial^{\lambda} \! \ln \Omega\right).
\label{transp}
\feqr

\noi In the general case (contortion and non-metricity tensor are included) the transformation law of 
the spin connection will be determined in the same way as in the Riemann case namely:
\beqr
\tilde{\omega}^{\,\,\, a}_{\mu \,\,\,\, b}=\omega^{\,\,\, a}_{\mu \,\,\,\, b}+ A^{\,\, \lambda}_{\mu \,\,\, \nu} 
e^{a}_{\lambda} e^{\nu}_{b}
\rightarrow \hat{\tilde{\omega}}^{\,\,\, a}_{\mu \,\,\,\, b}=
\hat{\omega}^{\,\,\, a}_{\mu \,\,\,\, b} + \hat{A}^{\,\, \lambda}_{\mu \,\,\, \nu} \hat{e}^{a}_{\lambda}
\hat{e}^{\nu}_{b}
\label{genspi} 
\feqr

\noi where $\hat{\omega}^{\,\,\, a}_{\mu \,\,\,\, b}$ is given by\rf{transp} and the vierbein postulate is suitably 
defined.

%%%%%%%%%%%%%%%%%%%%%%%%%%%%%%%%%%%%%%%%%%%%%%%%%%%%%%%%%%%%%%%%%%%%%%%%%%%%%%%%%%%%%%%%%%%%%%%%%%%%%%%%%%%%%%%%%%%%%%%

						%%%%%%%%%%%%%%%%%%
						%    SECTION 2   %
						%%%%%%%%%%%%%%%%%%

\section{Local Weyl invariance of the bosonic action.}

In n-dimensions ($n \neq 2$) the Einstein-Hilbert action with cosmological constant $\Lambda$ is:
\beqr
S_{EH}=\int d^n x \sqrt{g} \left(\frac{1}{G} R + \Lambda \right)
\label{einhi}
\feqr

\noi where $g=-\det g_{\mu \nu}$ and $G$ is Newton's constant. Under the local conformal transformations:
\beqr
\sqrt{g} &\rightarrow& \sqrt{\hat{g}} = \Omega^{\frac{n}{2}} \sqrt{g} \\
R &\rightarrow& \hat{R} = \Omega^{-1} \left[R - (n-1) D^{\mu} \partial_{\mu} \ln \Omega - \frac{1}{4}
(n-1)(n-2) \partial^{\mu} \ln \Omega \partial_{\mu} \ln \Omega \right]
\label{contrg}
\feqr

\noi it becomes:
\beqr
\hat{S}_{EH} = \int d^n x \sqrt{g} \, \Omega^{\frac{(n-2)}{2}} &\bigg[& \frac{1}{G} \left(R -  (n-1) D^{\mu} 
\partial_{\mu} \ln \Omega - \frac{1}{4} (n-1)(n-2) \partial^{\mu} \ln \Omega \partial_{\mu} \ln \Omega\right) 
\non \\
&+& \Lambda \Omega^{\frac{n}{2}}\bigg].
\label{tractw}
\feqr

\noi If one defines:
\beqr
\phi=\Omega^{\frac{n-2}{4}} \sqrt{\frac{2}{\xi G}}
\label{dephi}
\feqr

\noi with $\xi$ a dimensionless constant, to be determined later on, and partially integrate\rf{tractw} 
one arrives at:
\beqr
S_{EH}= \int d^n x \sqrt{g} \left[\frac{1}{2} g^{\mu \nu} \partial_{\mu}\phi \partial_{\nu}\phi + \frac{1}{2}
\xi R \phi^2 + \Lambda \left(\frac{\xi G}{2} \right)^{\frac{n}{n-2}} \phi^{\frac{2n}{n-2}}\right].
\label{nparde}
\feqr

\noi The first term in brackets is the kinetic part of the real scalar  
field  and the second for the distinguished value $\xi=\frac{1}{6}$ guarantees conformal invariance in the non-minimal 
coupling limit 
of $V_4$. As an exercise one can check that\rf{nparde} is indeed invariant under conformal rescalings of the 
metric\rf{conmet} provided that:
\beqr
\hat{\phi} &=& \Omega^{\frac{2-n}{4}} \phi \\
\xi &=& \frac{1}{4}\left(\frac{n-2}{n-1} \right).
\label{field} 
\feqr

Let us now consider scalar fields of conformal weight $d(\phi )$ i.e.
\beqr
\hat{\phi}=\Omega^{d (\phi )}\phi
\label{dfie}
\feqr 

\noi then we have:
\beq
\partial_{\mu} \hat{\phi}=\Omega^{d}\left(\partial_{\mu}+d(\phi )\partial_{\mu} \ln \Omega \right)\phi.
\feq{concov}

\noi One can replace $\partial_{\mu}$ with the conformally covariant derivative 
$\mathcal{D}_{\mu}=\partial_{\mu} - d(\phi )w_{\mu}$ transforming as:
\beq
\mathcal{D}_{\mu}\phi \rightarrow \Omega^{d(\phi)} \mathcal{D}_{\mu} \phi
\feq{covtra}

\noi where $w_{\mu} \rightarrow w_{\mu} + \partial_{\mu} \ln \Omega$ and $d(\mathcal{D}_{\mu}\phi)=d(\phi)$. 
For convenience we distinguish and study the following two cases separately.

\vspace{1.0 cm}
\textit{\bf Case I: Non-metricity tensor in $\bf (W_{n}, g)$ space-time.}

\vspace{0.5 cm}
In the presence of non-metricity tensor we construct the following massless bosonic action in n-dimensions:
\beqr
S=\frac{1}{2} \int \!\! \sqrt{g} \! \left[g^{\mu \nu} \mathcal{D}_{\mu} \phi \mathcal{D}_{\nu} \phi
+ \xi \mathcal{R}(\tilde{\Gamma}) \phi^2 + \Lambda \left(\frac{\xi G}{2} \right)^{\frac{n}{n-2}} \phi^{\frac{2n}{n-2}}
 \right] d^{n}x. 
\label{newact}
\feqr 

\noi For the $(W_n , g)$ space-time our affine connection is given by:
\beqr
\tilde{\Gamma}^{\lambda}_{\mu \nu}=\Gamma^{\lambda}_{\mu \nu} + \frac{1}{2} \left(g_{\mu \nu} w^{\lambda} -
\delta^{\lambda}_{\nu} w_{\mu} -  \delta^{\lambda}_{\mu} w_{\nu} \right).
\label{weylaff} 
\feqr 

\noi Expression\rf{weylaff} enjoys the following properties:
\beqr
\delta \tilde{\Gamma}^{\lambda}_{\mu \nu} &=& 0 \\
\tilde{\Gamma}^{\lambda}_{\mu \nu} &=& \tilde{\Gamma}^{\lambda}_{\nu \mu} \\
\tilde{\Gamma}^{\mu}_{\mu \nu} &=& \Gamma^{\mu}_{\mu \nu} - \frac{n}{2} w_{\mu} \\
g^{\mu \nu} \tilde{\Gamma}^{\lambda}_{\mu \nu} &=&  - \frac{1}{\sqrt{g}} \partial_{\mu}(\sqrt{g} g^{\mu \lambda})
+ \left(\frac{n-2}{2} \right) w^{\lambda}.
\label{proaff}
\feqr

\noi The first identity indicates the conformal invariance of the affine connection. The associated Ricci scalar has 
conformal weight $d(\mathcal{R}(\tilde{\Gamma}))=-1$ as one might check and the action is rewritten as:
\beqr
S = \frac{1}{2} \int d^n x \,\, \sqrt{g} \! &\bigg[& \!\!\! g^{\mu \nu} \mathcal{D}_{\mu} \phi \mathcal{D}_{\nu} \phi
 + \xi \left(R + (n-1) D^{\mu} w_{\mu} - \frac{1}{4} (n-1)(n-2) w^{\mu}w_{\mu} \right) \phi^2  \non \\ 
&+& \Lambda \left(\frac{\xi G}{2} \right)^{\frac{n}{n-2}} \phi^{\frac{2n}{n-2}} \bigg].
\label{rewact}
\feqr 

\noi The action\rf{rewact} is conformally invariant and when the Weyl vector field vanishes 
then $\mathcal{D}_{\mu} \phi = \partial_{\mu} \phi$ recovering\rf{nparde} in this limit. 

One might wonder if there exist additional terms that contribute to\rf{rewact} which do not  break local conformal 
invariance. To answer this question we classify first all $\textit{the building block tensors}$ according to 
their conformal weight which ranges in the integer interval $[-1, 2]$. 

\begin{table}[h] \caption{Building block tensors versus their conformal weights}
\begin{center} 
\begin{tabular}{|r|l|} \hline
\itshape Tensor & \itshape d(A) \\
\hline\hline
\makebox[1.5 cm] {$\Box,$ $\mathcal{R}$}                               &  \,-1 \\
\hline
\makebox[1.5 cm] {$\mathcal{R}_{\mu \nu}$}                             &   \, 0 \\
\hline
\makebox[1.5 cm] {$\mathcal{H}_{\mu \nu}$}                             &   \, 0 \\
\hline
\makebox[1.5 cm] {$g_{\mu \nu}, \mathcal{R}_{\mu \nu \lambda \rho}$}   &   \, 1 \\
\hline
\makebox[1.5 cm] {$\epsilon_{\mu \nu \lambda \rho}$}                   &   \, 2 \\
\hline
\end{tabular}
\end{center}
\end{table}

\noi In Table 1, $\Box = g^{\mu \nu} \mathcal{D}_{\mu} \mathcal{D}_{\nu}$ and  
$\mathcal{H}_{\mu \nu}(w)$ is the curvature of the free $w_{\mu}$ Weyl field defined in close analogy 
to electromagnetism 
as: $\mathcal{H}_{\mu \nu}= \mathcal{R}^{\rho}_{\rho \mu \nu}=2\mathcal{D}_{[\mu} w_{\nu ]}= 2\partial_{[\mu} w_{\nu ]} $. One might attempt 
to construct independent terms containing at most two $\mathcal{R}$'s and $\mathcal{H}$'s with total 
conformal weight -2 such as:

\beqr
\mathcal{R}_{\mu \nu}\mathcal{H}^{\mu \nu}, \quad g^{\mu \nu} \mathcal{R}_{\mu \nu \lambda \rho} \mathcal{R}^{\lambda \rho},
\quad \mathcal{R} \mathcal{R}_{\mu \nu \lambda \rho} \epsilon^{\mu \nu \lambda \rho}
\label{neadt}
\feqr
 
\noi but further inspection reveals that all of them fall into the same class since they are proportional 
to $\mathcal{H}_{\mu \nu} \mathcal{H}^{\mu \nu}$. 
The issue of the valiable additional terms will be presented in a more detail in section 5.

\vspace{1 cm}
\textit{\bf Case II: Contortion tensor in the $\bf (U_4, g)$ space-time.}

\vspace{0.5 cm}

In the absence of the traceless part $t_{\mu \nu \lambda}=0$ of the contortion the associated affine connection reads:
\beqr
\tilde{\Gamma}_{\mu \nu}^{\lambda}= \Gamma_{\mu \nu}^{\lambda}+ \frac{1}{3} \left(
\delta^{\lambda}_{\nu} v_{\mu} - \delta^{\lambda}_{\mu} v_{\nu}  + 6 \epsilon_{\mu \nu \ \ \alpha}^{\ \ \, \lambda}
S^{\alpha} \right)
\label{ctoaff} 
\feqr

\noi which together with  the scalar curvature:
\beqr
\mathcal{R}(K)= D^{\mu}v_{\mu} + \frac{1}{3}v^{2} + \frac{2}{3} S^{2}
\label{conrv}
\feqr

\noi are no longer conformally invariant when the vector part $v_{\mu}$ transforms like the Weyl field $w_{\mu}$ 
and the axial vector part $S_{\mu}$ has conformal weight $d(S_{\mu})=0$. 
%%%%%%%%%%%%%%%%%%%%%%%%%%%%%%%%%%%%%%%%%%%%%%%%%%%%%%%%%%%%%%%%%%%%%%%%%%%%%%%%%%%%%%%%%%%%%%%%%%%%%%%%%%%%%%%%%%%%%%%%%% 

					%%%%%%%%%%%%%%%%
					%   SECTION 3  %
					%%%%%%%%%%%%%%%%

\section{Local Weyl invariance of the fermionic spin-$1/2$ action.}

Our main purpose in this section is by exploiting the conformal invariance of the fermionic action under the 
local Weyl rescalings: 
\beqr
e^{\mu}_{a} \!&\rightarrow&\! \hat{e}^{\mu}_{a} = \Omega^{-1/2} e^{\mu}_{a} \\
e \!&\rightarrow&\! \hat{e} = \Omega^{2} e \\
\Psi \!&\rightarrow\!& \hat{\Psi} = \Omega^{-3/4} \Psi
\label{confmv}
\feqr

\noi to build an action that possibly contains the irreducible fields  
$\{t_{\mu \nu \lambda}, v_{\mu}, S_{\mu}, N_{\mu \nu \lambda}  \}$ and inherits conformal invariance. 

For this reason we consider the most general \footnote{In principle one could also add a gauge connection $A_{\mu}=
A^{\alpha}_{\mu} T_{\alpha}$ taking values in an arbitrary representation of an arbitrary Lie group but such a 
term is neglected at present but we will come to this point later.} complex (Dirac) spin-$\frac{1}{2}$ fermion action 
in four dimensions which is given by:
\beq
S_{1/2}=\int \! d^{4}x \, e \, \overline{\Psi} \, \nabla \!\!\!\! \slash \, \Psi,
\feq{action}

\noi where
\beqr
\nabla \!\!\!\!  \slash=e^{\mu}_{a}\gamma^{a} \left(\partial_{\mu}+ \frac{1}{4} \tilde{\omega}_{\mu mn}
\gamma^{m} \gamma^{n}\right)
\label{nab}
\feqr

\noi and
\beqr
\tilde{\omega}_{\mu mn}\gamma^{m} \gamma^{n}= \omega_{\mu mn}\gamma^{mn} + A_{\mu mn} 
\gamma^{m} \gamma^{n}.
\label{cotsp}
\feqr

\noi In\rf{nab} $e^{a}_{\mu}$ is the vierbein for the metric $g_{\mu \nu}$, $e=\det e^{a}_{\mu}$, $\gamma^{a}$ are the
$SO(4)$ Dirac matrices and $\gamma^{mn}=\frac{1}{2}[\gamma^{m},\gamma^{n}]$. 

\vspace{1 cm}
\textit{\bf Case I: Non-metricity tensor in $\bf (W_4, g)$ space-time.}

\vspace{0.5 cm}

We consider first the case in which the ``metric-compatibility'' condition  
(under parallel transport in a Weyl space a vector changes its magnitude as well as the relative rotation) is satisfied. 
The vierbein postulate now becomes:
\beqr
\left( \mathcal{D}_{\mu}+d(e_{a}^{\nu}) w_{\mu} \right)e^{a}_{\nu} =0 \quad \textit{or equivalently} \non \\
\nabla_{\mu}e^{a}_{\nu}=\partial_{\mu}e^{a}_{\nu}- \tilde{\Gamma}_{\mu \nu}^{\lambda} e^{a}_{\lambda} +
\tilde{\omega}_{\mu \ \ \ b}^{\ \ a} e^{b}_{\nu}= \frac{1}{2} w_{\mu} e^{a}_{\nu}
\label{nviep}
\feqr

\noi and the spin connection in terms of the vierbein is written as:
\beqr
\tilde{\omega}_{\mu a b} &=& \omega_{\mu ab} + {}_{w}\omega_{\mu ab} \non \\
&=& \frac{1}{2} \left(\lambda_{b \mu a}+\lambda_{\mu ab}+\lambda_{ba \mu} \right)+
w^{\nu} e_{\nu [a}e_{b] \mu}.
\label{spnmet}
\feqr

\noi Plugging into $(1/4) \gamma^{\mu} \omega_{\mu mn} \gamma^{m}\gamma^{n}$ 
the expression\rf{spnmet} and making use of the identities \rf{gid1} and\rf{gid2} of the Appendix, 
a straightforward computation gives the action:
\beqr
S_{1/2}=\int d^4 x e \bar{\Psi} e^{\mu}_{a} \gamma^{a} \left(D_{\mu} -\frac{3}{4}w_{\mu} \right)\Psi.
\label{acnmet}
\feqr

\noi Similarly to the bosonic case one could also define the conformal derivative 
$\mathcal{D}_{\mu}$ acting on the fermion field as:
\beqr
\mathcal{D}_{\mu}\Psi= \left(D_{\mu} - d(\Psi)w_{\mu}\right) \Psi
\label{confer} 
\feqr

\noi and then action\rf{acnmet} becomes invariant under local Weyl transformations. This can be proved  provided that 
the Weyl field transforms in the usual way $w_{\mu} \rightarrow w_{\mu} + \partial_{\mu} \ln \Omega$ and bearing in mind 
that $\gamma^{\mu} \delta \omega_{\mu mn}\gamma^m \gamma^n = -(n-1)\gamma^{\mu} \partial_{\mu} \ln \Omega$. The 
conclusion driven from the above analysis is that the fermionic action remains intact, when the covariant derivative 
is replaced by its appropriate conformal partner, under local Weyl transformations and the non-metricity tensor 
is present. In other words in a Weyl space the Dirac spinor does not couple to the Weyl vector $w_{\mu}$ although this 
does not preclude the existence of other types of coupling terms\cite{hochberg}.

\vspace{1 cm}
\textit{\bf Case II: Contortion tensor in $\bf (U_4, g)$ space-time.}

\vspace{0.5 cm}

Using\rf{conirr} one could evaluate the term:
\beqr
\frac{1}{4} \gamma^{\mu} K_{\mu mn} \gamma^{m}\gamma^{n}= \frac{3}{4} \gamma^{\mu} 
\left(\frac{2}{3}v_{\mu} -\gamma_{5} S_{\mu} \right). 
\label{relk}
\feqr
   
\noi It is worth noting that\rf{relk} is independent of the tensor $t_{\mu \nu \lambda}$ implying that
the fermions are only coupled to the vector and axial-vector parts of the torsion. 

The action then takes the form:
\beqr
S_{1/2}=\int \! d^{4}x \, e \, \overline{\Psi} \, e^{\mu}_{a} \, \gamma^{a}
\left[D_{\mu}+ \frac{1}{4}\Big( 2v_{\mu}-3\gamma_{5}S_{\mu} \Big) \right]\Psi
\label{actne}
\feqr

\noi Assuming that $d(S_{\mu})=0$ then apparently\rf{actne} can be written in an invariant way 
($v_{\mu}$ follows identical transformation law to that of the Weyl field) provided that $D_{\mu}$ 
is replaced by:
\beqr
\mathcal{D}_{\mu}\Psi=\left(D_{\mu} + \frac{2}{3}d(\Psi)v_{\mu}\right)\Psi.
\label{fecoco} 
\feqr

\noi From the final form of $S_{1/2}$ after performing the substitution\rf{fecoco} it is evident that the minimal Dirac 
action permits the spinor field to interact only with $S_{\mu}$ but not with $v_{\mu}$ and $t_{\mu \nu \lambda}$.

%%%%%%%%%%%%%%%%%%%%%%%%%%%%%%%%%%%%%%%%%%%%%%%%%%%%%%%%%%%%%%%%%%%%%%%%%%%%%%%%%%%%%%%%%%%%%%%%%%%%%%%%%%%%%%%%%%%%%%%%%%%%%

						%%%%%%%%%%%%%%%%%%%
						%    SECTION 4    %
						%%%%%%%%%%%%%%%%%%%

\section{Additional contributions}

Let us now examine what extra independent terms could possibly contribute to\rf{rewact} preserving conformal 
invariance of the free field theory and keeping at most quadratic terms in derivatives of fields and 
$\mathcal{R}$'s.
 
\begin{description}
\item[$1.$] \quad A mass term $\sqrt{g} m^{2}_{w} w^{\mu}w_{\mu}$ will impose a fixed length scale in the theory 
and thus it is excluded since local Weyl invariance will be softly broken. The appearance of such a term is 
expected if one desires to make explicit contact with the classical Einstein gravity \cite{hochberg}.

\item[$2.$] \quad A nonlinear $-\frac{\lambda}{4!}\sqrt{g} \phi^4$ potential does not break conformal invariance of 
the action but when we are interested in the free Langragian case it is neglected. 
An analogous quartic interaction with the 
scalar field $\phi$ replaced by $w_{\mu}$ does not pass the local Weyl transformation test as well.

\item[$3.$] \quad On the ground of low energy physics  higher derivative terms in\rf{rewact} with respect to $w_{\mu}$
are disregarded. The kinetic term of the Weyl field is proportional to the square of the field strength 
and a term of the form $(\partial_{\mu} w^{\mu})^2$ is abscent. This is explained by the fact that in a unitary vector 
theory both transversal and longitudinal components cannot propagate simultaneously \cite{odin}.
 
\item[$4.$] \quad The Euler form for an even dimensional manifold $\mathcal{M}$ is given by:
\beqr
e_{2l}(\mathcal{M}) &=& \frac{(-1)^l}{(4\pi)^l l!} \epsilon^{i_1 i_2 \cdots i_{2l}} R_{i_1 i_2} \wedge \cdots 
R_{i_{2l-1} i_{2l}} \non \\
&=& \frac{(-1)^l}{(4\pi)^l l!} I_{2l} \ \textit{dVol}
\label{eulfor}  
\feqr

\noi where $R^{i}_{\ \, j}=\frac{1}{2} R^{i}_{\ \, jkl} e^{k}_{\mu} e^{l}_{\nu} dx^{\mu} \wedge dx^{\nu}$, $I_{2l}= 2^{-l} 
\epsilon^{i_1 i_2 \cdots i_{2l}} \epsilon^{j_1 j_2 \cdots j_{2l}} R_{i_1 i_2 j_1 j_2} \cdots R_{i_{2l-1}i_{2l}j_{2l-1}j_{2l}}$
 and $\textit{dVol}=\sqrt{g} \, d^{2l}x$ is the invariant 
volume element. Integration of $e_{2l}(\mathcal{M})$ over a compact orientable Riemannian manifold 
provides the Gauss-Bonnet theorem expressed by:
\beqr
\chi_{2l}(\mathcal{M})=\int_{\mathcal{M}_{2l}} e_{2l}(\mathcal{M}).
\label{gausbo} 
\feqr

\noi In four dimensions $I_4 = R_{ijkl}R^{ijkl}-4R_{ij}R^{ij}+R^2$ and the Gauss-Bonnet theorem in a coordinate basis 
becomes:
\beqr
\chi_{4}(\mathcal{M})=\frac{1}{32 \pi ^2} \int_{\mathcal{M}_4} \sqrt{g} \left( R_{\mu \nu \lambda \rho} 
R^{\mu \nu \lambda \rho} -4R_{\mu \nu} R^{\mu \nu} + R^2\right)d^4 x 
\label{gb4} 
\feqr
 
\noi  The topological action\rf{gb4} is also 
conformally invariant since $d(R_{\mu \nu \lambda \rho})=1$ and $d(R_{\mu \nu})=0$. 

\item[$5.$] \quad Constructing the dualized tensor:
\beqr
{}^{\ast}R_{\mu \nu \lambda \rho} &=& \frac{1}{2} \epsilon_{\mu \nu \sigma \tau} 
R^{\sigma \tau}_{\ \ \  \lambda \rho} 
\label{duat}
\feqr

\noi then the Pontrjagin number in four dimensions reads:
\beqr
\mathcal{P}_4 &=& \frac{1}{8\pi^2}\int_{\mathcal{M}_4} R^{ij}\wedge R_{ij} \non \\
&=& \frac{1}{16\pi^2} \int_{\mathcal{M}_4} \!\! \sqrt{g} \,\, {}^{\ast} \check{R}^{\mu \nu \lambda \rho}
\check{R}_{\mu \nu \lambda \rho} \, \, d^4 x 
\label{ponn}
\feqr
 
\noi  This parity odd topological quantity is conformally invariant as one might check with the help of
 $d({}^{\ast}R_{\mu \nu \lambda \rho})=1$.  

\item[$7.$] \quad Another topological invariant is the winding number:
\beqr
N_4 = \frac{1}{32 \pi^2}\int d^4 x \sqrt{g} \,\, {}^{\ast}\mathcal{H}_{\mu \nu} \mathcal{H}^{\mu \nu}
\label{winn}
\feqr

\noi where ${}^{\ast}\mathcal{H}_{\mu \nu}$ is the dualized field strength of the associated field.

\item[$8.$] \quad When the torsion is present there is a topological invariant provided by the Nieh-Yan 
locally exact four form \cite{nieya}:
\beqr
N_4 &=& \int_{\mathcal{M}_4} \left(T^i \wedge T_i - R_{ij}\wedge \theta^{i} \wedge \theta^{j} \right) \non \\
&=& \frac{1}{2} \int \sqrt{g} \left(\frac{1}{2} T^{\sigma}_{\ \, \mu \nu} T_{\sigma \lambda \rho} - 
R_{\mu \nu \lambda \rho}\right) \epsilon^{\mu \nu \lambda \rho} \, d^4 x
\label{nieh} 
\feqr

\noi where $\theta^{i}=e^{i}_{\mu} dx^{\mu}$ is a non-coordinate basis. This invariant is not conformally 
inavariant since 
the N-Y form is a function of the local frame as opposed to the topological invariants discussed above.

\item[$9.$] \quad The term $\sqrt{g}R^2$ is permissible and actually needed as a counterterm to the ``two point'' 
infinity. The term $\sqrt{g}\ \Box R = \sqrt{g} g^{\mu \nu} \mathcal{D}_{\mu} \mathcal{D}_{\nu} R$ is also acceptable.

\item[$10.$] \quad A Yakawa coupling of the form $\sqrt{g}\phi \bar{\Psi} \Psi$ is acceptable and generates mass 
to the fermions when present.

\item[$11.$] \quad In $(W_n, g)$ ($n \geq 3 $) space-time there is also the Weyl tensor which is conformally 
invariant and given by:
\beqr
C_{\mu \alpha \beta \gamma}(\Gamma, M) &=& R_{\mu \alpha \beta \gamma}(\Gamma) - \frac{1}{n-2} 
\left(g_{\mu \beta} R_{\alpha \gamma} + g_{\alpha \gamma} R_{\mu \beta} - g_{\mu \gamma} R_{\alpha \beta} 
- g_{\alpha \beta} R_{\mu \gamma} \right)(\Gamma) \non \\ 
&-& \frac{1}{(n-1)(n-2)} R(\Gamma) 
\left(g_{\mu \gamma} g_{\alpha \beta} - g_{\mu \beta} g_{\alpha \gamma}\right) + X_{\mu \alpha \beta \gamma}
(\mathcal{H})
\label{WeyW} 
\feqr

\noi where $g^{\alpha \gamma} C_{\mu \alpha \beta \gamma} = 0$, 
\beqr
X_{\mu \alpha \beta \gamma}(\mathcal{H}) = \frac{1}{2(n-2)} \left(g_{\mu \beta} \mathcal{H}_{\alpha \gamma} +
g_{\alpha \gamma} \mathcal{H}_{\mu \beta} - g_{\mu \gamma} \mathcal{H}_{\alpha \beta} - 
g_{\alpha \beta} \mathcal{H}_{\mu \gamma}\right) + \frac{1}{2} g_{\mu \alpha} \mathcal{H}_{\beta \gamma}(w) 
\label{Xabc}
\feqr

\noi and
\beqr
C^{2}_{\mu \alpha \beta \gamma}(\Gamma, M) &=& R^{2}_{\mu \alpha \beta \gamma}(\Gamma) 
- \frac{4}{n-2} R^{2}_{\alpha \beta} (\Gamma) + \frac{2}{(n-1)(n-2)} R^2 (\Gamma) \non \\
&+& \frac{(n^2 - 2n +4)}{4(n-2)} \mathcal{H}^{2}_{\alpha \beta}.
\label{sqww}
\feqr

\noi So we observe from\rf{sqww} that the field strength could be inserted into the gravity action in a natural 
way avoiding addition of such a term by force. 

The situation is drastically changed if one moves on to a subgroup of the $(Y_4, g)$ space-time and considers the existence of 
torsion with only totally antisymmetric part:
\beqr
K_{\mu \nu \lambda}=\frac{1}{2} \epsilon_{\mu \nu \lambda \rho} S^{\rho} 
\label{toanti}
\feqr

\noi and $d(K_{\mu \nu \lambda})= 1$. A cumbersome but straightforward calculation gives the extra tensor:
\beqr
X_{\mu \alpha \beta \gamma}(S) &=& - \frac{1}{2} \bigg[ g_{\mu [ \beta}g_{\gamma ] \alpha}S^2 + 
2 \epsilon_{\mu \alpha \rho [ \gamma} D_{\beta ]} S^{\rho} + 
\left(g_{\mu [ \beta} \epsilon_{\gamma ] \rho \alpha \sigma} - 
g_{\alpha [ \beta} \epsilon_{\gamma \rho \mu \sigma}\right) \non \\
&+& \epsilon_{\mu \sigma \rho [ \beta} \epsilon_{\gamma ] \,\,\,\, \alpha \lambda}^{\,\,\,\, \sigma} 
S^{\rho} S^{\lambda} - \left(g_{\mu [ \beta} S_{\gamma ]} S_{\alpha} -
g_{\alpha [ \beta} S_{\gamma ]} S_{\mu} \right) \bigg].
\label{exten}
\feqr  

\noi The term $\epsilon_{\mu \alpha \rho [ \gamma}D_{\beta ]} S^{\rho}$ in\rf{exten} 
spoils conformal invariance but one has the freedom to define the following conformally invariant tensor: 
\beqr
Y_{\mu \alpha \beta \gamma}(S) = X_{\mu \alpha \beta \gamma}(S) + 
\epsilon_{\mu \alpha \rho [ \gamma} D_{\beta ]} S^{\rho}.
\label{defty}  
\feqr

\noi and as a consequence the quadratic \textit{Weyl-like} tensor  
\beqr
C^{2}_{\mu \alpha \beta \gamma}(\Gamma, M, S) &=& R^{2}_{\mu \alpha \beta \gamma}(\Gamma) 
- \frac{1}{2} R^{2}_{\alpha \beta}(\Gamma) + \frac{1}{3}R^2 (\Gamma) \non \\
&+& \frac{3}{2} \left(\mathcal{H}^{2}_{\alpha \beta}(w) + (S^2)^2 \right) - 
2D^{\alpha}S^{\beta} \mathcal{H}_{\alpha \beta}(S)
\label{wesqr}
\feqr

\noi is conformally invariant. Notice here the appearance of the extra coupling terms 
$\mathcal{H}^{2}_{\mu \nu}(w)$, $(S^2)^2$ and $D^{\alpha} S^{\beta} \mathcal{H}_{\alpha \beta}(S)$.

\item[$12.$] \quad Finally we address the possibility the Weyl strength $\mathcal{H}_{\mu \nu}$ to interact with the 
Yang-Mills field strength. In the non-Abelian case one could try the following coupling:
\beqr
\alpha tr\left(\mathcal{H}^{\mu \nu} \mathcal{F}_{\mu \nu} \right)
\label{nabel}
\feqr 

\noi where $\mathcal{F}_{\mu \nu}=T^{a}F^{a}_{\mu \nu}$, $a=1, \cdots , dim(\mathcal{G})$. The Langrangian\rf{nabel} is 
locally scale invariant but not gauge invariant under the action of the symmetry group $\mathcal{G}$. The situation 
is improved in the Abelian case with electromagnetic field strength $F_{\mu \nu}$ ($U(1)$ or F-curvature). 
Admissible candidates are:
\beqr
\beta_{1} \mathcal{H}^{\mu \nu} F_{\mu \nu}; \quad  
\beta_{2} \epsilon^{\mu \nu \rho \sigma} \mathcal{H}_{\mu \nu} F_{\rho \sigma} 
\label{abel}
\feqr

\noi with $d(A_{\mu})=d(F_{\mu \nu})=0$. One might convince itself that these types of interactions are simultaneously
locally Weyl and $U(1)$ invariant.   
\end{description}

%%%%%%%%%%%%%%%%%%%%%%%%%%%%%%%%%%%%%%%%%%%%%%%%%%%%%%%%%%%%%%%%%%%%%%%%%%%%%%%%%%%%%%%%%%%%%%%%%%%%%%%%%%%%%%%%%%%%%

						%%%%%%%%%%%%%%%%%%%%%
						%    CONCLUSIONS    %
						%%%%%%%%%%%%%%%%%%%%%

\section{Conclusions}

The lessons this exercise taught us are summarized in the following:

\begin{description}

\item[$A.$] \quad The Einstein-Hilbert action can be conformally reduced to a massless bosonic action in n-dimensions 
and the non-metricity and torsion tensor can be 
included in a conformally invariant way respectively. This is achieved by defining the conformal partner of the 
covariant derivative and adjusting the affine connection in the case of torsion.

\item[$B.$] \quad A conformally invariant fermionic action for the spin-$1/2$ field is also constructed for the 
two cases under consideration. Applying the idea of conformal derivative we write down the corresponding actions and
show that the vector and tensorial parts of the Cartan's torsion do not couple minimally to fermions. 

\item[$C.$] \quad There are certain topological invariants such as the Euler characteristic, the Pontrjagin and 
winding numbers that contribute to the action without affecting the conformal properties of the theory. 
In a subgroup of the $(Y_4, g)$ space-time one could also construct a  \textit{Weyl-like} tensor the square of which 
provides new terms to the gravity action such as $\mathcal{H}^{2}_{\mu \nu}(w)$, $(S^2)^2$ and 
$D^{\alpha}S^{\beta} \mathcal{H}_{\alpha \beta}(S)$. We also examine the possibility of coupling the photons to 
the Weyl vector.

\end{description} 

%%%%%%%%%%%%%%%%%%%%%%%%%%%%%%%%%%%%%%%%%%%%%%%%%%%%%%%%%%%%%%%%%%%%%%%%%%%%%%%%%%%%%%%%%%%%%%%%%%%%%%%%%%%%%%%%%%%%%%%%%

					%%%%%%%%%%%%%%%%%%
					%ACKNOWLEDGEMENTS%
					%%%%%%%%%%%%%%%%%%

\section*{Acknowledgements}

I am grateful to F. W. Hehl for suggesting some crucial improvements to the manuscript and also 
to thank T. Christodoulaki and G. Diamandi for valuable discussions.

%%%%%%%%%%%%%%%%%%%%%%%%%%%%%%%%%%%%%%%%%%%%%%%%%%%%%%%%%%%%%%%%%%%%%%%%%%%%%%%%%%%%%%%%%%%%%%%%%%%%%%%%%%%%

					%%%%%%%%%%%%%%%%%
					%   APPENDICES  %
					%%%%%%%%%%%%%%%%%

\vspace{0.5 cm}

\appendix
\section*{Appendix}
\setcounter{section} {1}
\setcounter{equation} {0}
\indent

Throughout this paper flat (tangent space) indices are denoted by Latin letters subjected
to local Lorentz rotations and boosts. Greek characters stand for curved indices and are
only subjected to local translations. Roman indices may be raised using the flat-space metric
$\eta^{ab}$; and Greek indices may be raised or lowered using:
\beq
g^{\mu\nu}(x)=\eta^{ab} e^{\mu}_{a}(x) e^{\nu}_{b}(x).
\feq{vielbe}

\noi Anti-symmmetrisations carry a weight of $1/n!$ where $n$ is the number of indices involved in these
operations.
 
The symbols we reserve to denote different types of covariant derivatives are:
\beqr
D_{\mu} && \quad \textit{for covariant derivative based on the Christoffel symbol}, \\
\mathcal{D}_{\mu} && \quad \textit{for conformal convariant derivative and} \\
\nabla_{\mu} && \quad \textit{for covariant derivative built on an affine connection} 
\label{deficov}
\feqr

The covariant derivative of a tensor with q-contravariant and p-covariant indices ($p > 1$) is defined as:
\beq
D_{\mu}T^{\alpha_1 \alpha_2 \cdots \alpha_q}_{\beta_1 \beta_2 \cdots \beta_p} = \partial_{\mu}
T^{\alpha_1 \alpha_2 \cdots \alpha_q}_{\beta_1 \beta_2 \cdots \beta_p} + 
\left[\Gamma_{\mu}, T \right]^{\alpha_1 \alpha_2 \cdots \alpha_q}_{\beta_1 \beta_2 \cdots \beta_p}
\feq{deften}

\noi  where $\left[\Gamma_{\mu}, T \right]^{\alpha_1 \alpha_2 \cdots \alpha_q}_{\beta_1 \beta_2 \cdots \beta_p}
=\Gamma_{\mu \, \nu}^{ \ \ \ \alpha_1} \, T^{\nu \alpha_2 \cdots \alpha_q}_{\beta_1 \beta_2 \cdots \beta_p}
+ \cdots + \Gamma_{\mu \, \nu}^{ \ \ \ \alpha_q} \, T^{\alpha_1  \alpha_2 \cdots \nu}_{\beta_1 \beta_2 \cdots \beta_p}
- (-1)^p \Gamma_{\mu \, \beta_1}^{ \ \ \ \nu} \, T^{\alpha_1 \alpha_2 \cdots \alpha_q}_{\nu \beta_2 \cdots \beta_p}
- \cdots - $ \\ $-(-1)^p \Gamma_{\mu \, \beta_p}^{ \ \ \ \nu} 
\, T^{\alpha_1 \alpha_2 \cdots \alpha_q}_{\beta_1 \beta_2 \cdots \nu}$, $p \geq 2$ and $\Gamma_{\mu \nu}^{\ \ \ \lambda}
=\{ {\lambda \atop\mu\nu} \}= \frac{1}{2} g^{\lambda \rho}\left(\partial_{\mu} g_{\nu \rho} + \partial_{\nu} g_{\mu \rho}
- \partial_{\rho} g_{\mu \nu} \right)$
is the usual Christoffel symbol. The commutation relation of the covariant derivatives is:

\beq
\left[\nabla_{\mu}, \nabla_{\nu} \right] V_{\lambda} = - R_{\mu \nu \lambda}^{\,\,\,\,\,\,\,\,\, \rho} \, V_{\rho} - 
\left[ 2 K_{\mu \nu}^{\,\,\,\,\,\, \rho} + N_{\mu \nu}^{\,\,\,\,\,\, \rho} - N_{\nu \mu}^{\,\,\,\,\,\, \rho}
\right] \nabla_{\rho} V_{\lambda}.
\feq{defcov}

One could also define a conformally covariant derivative for a tensor of the general type
$T^{\alpha_1 \alpha_2 \cdots \alpha_q}_{\beta_1 \beta_2 \cdots \beta_p}$ having conformal weight
$d(T^{\alpha_1 \alpha_2 \cdots \alpha_q}_{\beta_1 \beta_2 \cdots \beta_p})=d(T)$ as follows:   
\beqr
\mathcal{D}_{\mu}T^{\alpha_1 \alpha_2 \cdots \alpha_q}_{\beta_1 \beta_2 \cdots \beta_p}=
\nabla_{\mu}T^{\alpha_1 \alpha_2 \cdots \alpha_q}_{\beta_1 \beta_2 \cdots \beta_p} -
d(T) w_{\mu} T^{\alpha_1 \alpha_2 \cdots \alpha_q}_{\beta_1 \beta_2 \cdots \beta_p}
\label{deficonf}
\feqr

\noi where $w_{\mu}$ is the Weyl field. Thus for example the conformal derivative of scalar field is: 
$\mathcal{D}_{\mu} \phi = \partial_{\mu}- d(\phi ) w_{\mu} \phi$ while for the metricity tensor 
$g_{\mu \nu}$ with conformal weight $d(g_{\mu \nu})=1$ holds: $\mathcal{D}_{\mu} g_{\nu \lambda}=0$.   
The covariant derivative of a vector $A_{\mu}$ is given by:

\beqr
\nabla^{\mu} A_{\mu}=D^{\mu} A_{\mu} - A_{\mu} w^{\mu}=\nabla_{\mu} A^{\mu}+ A_{\mu} w^{\mu}.
\label{defivec}
\feqr

\noi The commutation of the conformal derivatives gives:

\beq
[\mathcal{D}_{\mu}, \mathcal{D}_{\nu}]A_{\lambda}=-\left(R_{\mu \nu \lambda}{}^{\rho} A_{\rho}
+ d(A_{\lambda})\mathcal{H}_{\mu \nu} A_{\lambda} \right).
\feq{concomu}

The Riemann curvature, Ricci tensor and the scalar curvature are defined by:
\beqr
R^{\mu}_{\,\,\, \rho \sigma \nu} &=& \partial_{\rho} \Gamma^{\mu}_{\,\,\, \sigma \nu} + \Gamma^{\mu}_{\,\,\, \rho \alpha}
\Gamma^{\alpha}_{\,\,\, \sigma \nu} - \rho \leftrightarrow \sigma \\
R_{\rho \nu} &=& R^{\mu}_{\,\,\, \rho \mu \nu} \\
R &=& g^{\mu \nu} R_{\mu \nu}
\label{defirr}
\feqr

\noi and in the presence of torsion and non-metricity tensors (with $L_{\mu \nu \lambda} \neq 0$) are given by:

\beqr
\mathcal{R}^{\mu}_{\,\, \nu \lambda \rho} (\Gamma, A) &=& \mathcal{R}^{\mu}_{\,\, \nu \lambda \rho}(\Gamma, M) +
\mathcal{R}^{\mu}_{\,\, \nu \lambda \rho}(K) - w^{\sigma} K^{\mu}_{\,\, \sigma [ \rho} \delta^{\nu}_{\lambda ]} +
w^{\mu} K^{\sigma \nu}_{\,\,\, [\rho} g_{\lambda ] \sigma} 
- w_{\sigma} K^{\sigma \nu}_{\,\,\, [\rho} \delta^{\mu}_{\lambda ]} \non \\
&-& w^{\nu} K^{\mu}_{\,\, [\rho \lambda]} - 
K^{\mu}_{\,\, \sigma [\rho} \mathcal{P}^{\sigma}_{\,\, \lambda] \nu} 
- K^{\sigma}_{\,\, \nu [\rho} \mathcal{P}^{\mu}_{\,\, \lambda] \sigma}
\label{riecur1}
\feqr

\beqr 
\mathcal{R}_{\nu \rho}(\Gamma, A) &=& \mathcal{R}_{\nu \rho}(\Gamma, M) + \mathcal{R}_{\nu \rho}(K) \non \\
&-& \frac{1}{2} \left[g_{\nu \rho} w^{\mu}v_{\mu} - v_{\rho}w_{\nu} +w^{\mu} (2K_{\mu \nu \rho} 
+ K_{\nu \mu \rho} - K_{\rho \nu \mu}) \right] \non \\
&+& \frac{1}{2} \left[ v_{\sigma} \mathcal{P}^{\sigma}_{\,\,\, \rho \nu} + K^{\mu}_{\,\,\, \sigma \rho} 
\mathcal{P}^{\sigma}_{\,\,\, \mu \nu} -
K^{\sigma}_{\,\,\, \nu \mu} \mathcal{P}^{\mu}_{\,\,\, \rho \sigma}\right]
\label{riecur2}
\feqr

\beqr
\mathcal{R}(\Gamma, A) = \mathcal{R}(\Gamma, M) + \mathcal{R}(K) - \frac{1}{2}
\left[5w^{\mu}v_{\mu} - v^2 + v_{\mu}\mathcal{P}^{\mu \,\,\,\,\, \nu}_{\,\,\, \nu} + K^{\mu}_{\,\,\, \sigma \nu}
(\mathcal{P}^{\sigma \,\,\,\,\, \nu}_{\,\,\, \mu}-\mathcal{P}^{\nu \sigma}_{\,\,\,\,\, \mu}) \right], 
\label{riecur3}
\feqr

\noi where:
\beqr
\mathcal{R}^{\mu}_{\,\,\, \nu \lambda \rho}(\Gamma, M) &=& R^{\mu}_{\,\,\, \nu \lambda \rho}(\Gamma) -
g_{\nu [ \lambda} D_{\rho ]}w^{\mu} + \delta^{\mu}_{\nu} D_{[\rho} w_{\lambda ]} 
+ \delta^{\mu}_{[ \lambda}D_{\rho]}w_{\nu} \non \\
&-& \frac{1}{2} \left[g_{\nu [ \lambda}w_{\rho ]}w^{\mu} + \delta^{\mu}_{[\lambda}g_{\rho ] \nu}w^2 +
\delta^{\mu}_{[\rho}w_{\lambda ]}w_{\nu} \right] \non \\
&-& D_{[\rho} \mathcal{P}^{\mu}_{\,\,\, \lambda ] \nu} - \frac{1}{4} w^{\mu} \left(\mathcal{P}_{\rho \nu \lambda} 
- \mathcal{P}_{\lambda \nu \rho}\right) \non \\
&-& \frac{1}{2} \left[w_{\sigma} \delta^{\mu}_{[\lambda} \mathcal{P}^{\sigma}_{\,\,\, \rho ] \nu} +
w^{\sigma}g_{\nu [ \lambda} \mathcal{P}^{\mu}_{\,\,\, \rho ] \sigma}
 + \mathcal{P}^{\mu}_{\,\,\, \sigma [ \rho} \mathcal{P}^{\sigma}_{\,\,\, \lambda ] \nu}\right]
\label{riemetr1}
\feqr

\beqr
\mathcal{R}_{\nu \rho}(\Gamma, M) &=& R_{\nu \rho} + \frac{1}{2} \left( g_{\nu \rho} D_{\mu} w^{\mu} 
+ 3D_{\rho}w_{\nu}
- D_{\nu}w_{\rho} - g_{\nu \rho} w^2 + w_{\nu}w_{\rho} \right) \non \\
&+& \frac{1}{2}D_{\mu} \mathcal{P}^{\mu}_{\,\,\, \rho \nu} - \frac{1}{4}w^{\mu}(2\mathcal{P}_{\mu \rho \nu} +
\mathcal{P}_{\rho \nu \mu} + \mathcal{P}_{\nu \rho \mu}) - \frac{1}{4} \mathcal{P}^{\mu}_{\,\,\, \sigma \rho}
\mathcal{P}^{\sigma}_{\,\,\, \mu \nu} 
\label{riemetr2}
\feqr

\beqr
\mathcal{R}(\Gamma, M) = R + 3D_{\mu}w^{\mu} - \frac{3}{2} w^2 + \frac{1}{2}
D_{\mu}\mathcal{P}^{\mu \nu}_{\,\,\,\,\, \nu} - \frac{1}{4}
(2w^{\mu} \mathcal{P}^{\,\,\,\,\, \nu}_{\mu \nu} + \mathcal{P}^{\mu}_{\,\,\, \sigma \nu} 
\mathcal{P}^{\sigma \nu}_{\,\,\,\,\, \mu})
\label{riemetr3}
\feqr

\beqr
\mathcal{R}^{\mu}_{\,\,\, \nu \lambda \rho}(K) &=& D_{\lambda} K^{\mu}_{\,\,\, \nu \rho} - 
D_{\rho} K^{\mu}_{\,\,\, \nu \lambda} - K^{\mu}_{\,\,\, \sigma \rho} K^{\sigma}_{\,\,\, \nu \lambda}
+ K^{\mu}_{\,\,\, \sigma \lambda} K^{\sigma}_{\,\,\, \nu \rho}
\label{rietor1}
\feqr 

\beqr
\mathcal{R}_{\nu \rho}(K) &=& D^{\mu} K_{\mu \nu \rho} + D_{\rho}v_{\nu} - v_{\sigma}
K^{\sigma}_{\,\,\, \nu \rho} - K^{\mu}_{\,\,\, \sigma \rho}K^{\sigma}_{\,\,\, \nu \mu} 
\label{rietor2} 
\feqr

\beqr
\mathcal{R}(K) = 2D_{\mu}v^{\mu} - v^2 - K^{\mu \,\,\,\,\, \nu}_{\,\,\, \sigma} K^{\sigma}_{\,\,\, \nu \mu},
\label{rietor3}
\feqr

\noi $R^{\mu}_{\,\,\, \nu \lambda \rho}(\Gamma)$ is the Riemann curvature constructed out of the 
Christoffel symbol $\Gamma^{\,\,\, \mu}_{\nu \lambda}$  and 
$\mathcal{P}_{\mu \nu \lambda}= L_{\mu \nu \lambda} - L_{\lambda \mu \nu} - L_{\nu \mu \lambda}$. They 
satisfy the following symmetries and identities:

\begin{description}

\item [\textit{Case I:} $\mathcal{R}^{\mu}_{\,\,\, \nu \lambda \rho}(\Gamma, M)$]

\beqr
\mathcal{R}^{\mu}_{\,\,\, \nu \lambda \rho} &=& - \mathcal{R}^{\mu}_{\,\,\, \nu \rho \lambda} \\
\mathcal{R}^{\mu}_{\,\,\, \{ \nu \lambda \rho \} } &=& 0 \\ 
\mathcal{D}_{\sigma} \mathcal{R}_{\mu \nu \lambda \rho} + \mathcal{D}_{\rho} \mathcal{R}_{\mu \nu \sigma \lambda}
+ \mathcal{D}_{\lambda} \mathcal{R}_{\mu \nu \rho \sigma} &=& 0 \\
\mathcal{D}_{ \{ \mu} \mathcal{H}_{ \nu \lambda \}} &=& 0 \\
\mathcal{R}_{\nu \lambda} - \mathcal{R}_{\lambda \nu} &=& 2\mathcal{H}_{\lambda \nu}  
\label{syM1}
\feqr

\noi where $\mathcal{H}_{\nu \lambda}=\partial_{\nu} w_{\lambda} - \partial_{\lambda} w_{\nu}$ and $\{ \mu \nu \lambda\}$ 
denotes cyclic permutation of the indices.

\item [\textit{Case II:} $\mathcal{R}^{\mu}_{\,\,\, \nu \lambda \rho}(K)$]
\beq
\mathcal{R}^{\mu}_{\,\,\, \nu \lambda \rho} = - \mathcal{R}^{\mu}_{\,\,\, \nu \rho \lambda}
\feq{syK1}

\beq
\mathcal{R}_{\nu \lambda} - \mathcal{R}_{\lambda \nu} = 2 \left[ \mathcal{H}_{[\nu \lambda]} 
- D_{\mu} K^{\mu}_{\ \ [\nu \lambda ]}
+ v_{\sigma}K^{\sigma}_{\ \ [\nu \lambda ]} + K^{\mu}_{\ \ \sigma [\lambda} K^{\sigma}_{\ \ \nu ]\mu}\right].
\feq{syK2}

\end{description}

In $(W_n, g)$ space-time the Ricci tensor is:
\beqr
\mathcal{R}_{\mu \nu}(\Gamma, M)= R_{\mu \nu}(\Gamma) + \left[g_{\mu \nu} D^{\alpha} w_{\alpha}
+ (n-2) D_{\nu}w_{\mu} \right] - \frac{1}{4}(n-2) (g_{\mu \nu} w^2 - w_{\mu} w_{\nu}) 
- \frac{1}{2} \mathcal{H}_{\mu \nu} 
\label{riwn}
\feqr

\noi while the Ricci scalar is given by:
\beqr
\mathcal{R}(\Gamma, M) = R(\Gamma) + (n-1) D^{\alpha} w_{\alpha} - \frac{1}{4} (n-1)(n-2) w^2.
\label{scalwn}
\feqr

Consider manifolds with Euclidean signature and hermitian Dirac matrices satisfying:

\beq
\{\gamma^{a},\gamma^{b}\}=2\delta^{ab},
\feq{gcom}

\noi $(\gamma^{a})^2=I$ and $(\gamma^{a})^{\dagger}=\gamma^{a}$, where $a=1,\cdots,5$. In 
this representation we make use of the following gamma matrix identities:

\beqr
\gamma_{a}\gamma_{b}\gamma_{c}=\gamma_{a}\delta_{bc}-\gamma_{b}\delta_{ac}+\gamma_{c}
\delta_{ab}+\epsilon_{abcd}\gamma_{5}\gamma_{d}
\label{gid1}
\feqr

\beqr
[ \gamma_{a},\gamma_{b} ]=-\epsilon_{abcd}\gamma_{5}\gamma^{cd}.
\label{gid2}
\feqr

\bibliographystyle{plain}

\end{document}